\documentclass[10pt,conference]{IEEEtran}
\IEEEoverridecommandlockouts

\usepackage{balance} \usepackage{cite}
\usepackage{amsthm} \usepackage{amsmath,amssymb,amsfonts}
\usepackage{algorithmic}
\usepackage{graphicx}
\usepackage{textcomp}
\usepackage{xcolor} \usepackage{booktabs}
\usepackage[flushleft]{threeparttable}
\usepackage{multirow}
\usepackage{bm}
\usepackage{tabu}
\usepackage{framed}
\usepackage{url}
\def\BibTeX{{\rm B\kern-.05em{\sc i\kern-.025em b}\kern-.08em
    T\kern-.1667em\lower.7ex\hbox{E}\kern-.125emX}}

\usepackage{enumitem}  \definecolor{SithColor}{rgb}{0.7,0,0}
\definecolor{ConsularColor}{rgb}{0,0.4,0} \definecolor{GuardianColor}{rgb}{0,0,0.8} \definecolor{WarningColor}{rgb}{255,97,0}

\newcommand{\ra}[1]{\renewcommand{\arraystretch}{#1}}
\newcommand{\tworow}[1]{\multirow{2}{*}[-3pt]{#1}}
\newcommand{\tabincell}[2]{\begin{tabular}{@{}#1@{}}#2\end{tabular}}

\begin{document}
\bstctlcite{bstctl:nodash}

\title{Automatic Generation of Pull Request Descriptions}

\author{
\IEEEauthorblockN{Zhongxin Liu\IEEEauthorrefmark{1}\IEEEauthorrefmark{2}\IEEEauthorrefmark{6}, Xin Xia\IEEEauthorrefmark{3}\textsuperscript{\checkmark}\thanks{\textsuperscript{\checkmark}Corresponding author.}, Christoph Treude\IEEEauthorrefmark{4}, David Lo\IEEEauthorrefmark{5}, Shanping Li\IEEEauthorrefmark{1}}
\IEEEauthorblockA{\IEEEauthorrefmark{1}College of Computer Science and Technology, Zhejiang University, Hangzhou, China}
\IEEEauthorblockA{\IEEEauthorrefmark{2}Ningbo Research Institute, Zhejiang University, Ningbo, China}
\IEEEauthorblockA{\IEEEauthorrefmark{6}PengCheng Laboratory, Shenzhen, China}
\IEEEauthorblockA{\IEEEauthorrefmark{3}Faculty of Information Technology, Monash University, Melbourne, Australia}
\IEEEauthorblockA{\IEEEauthorrefmark{4}School of Computer Science, University of Adelaide, Adelaide, Australia}
\IEEEauthorblockA{\IEEEauthorrefmark{5}School of Information Systems, Singapore Management University, Singapore, Singapore}
\IEEEauthorblockA{liu\_zx@zju.edu.cn, xin.xia@monash.edu, christoph.treude@adelaide.edu.au, davidlo@smu.edu.sg, shan@zju.edu.cn}
}

\maketitle

\begin{abstract}
Enabled by the pull-based development model, developers can easily contribute to a project through pull requests (PRs). When creating a PR, developers can add a free-form description to describe \emph{what} changes are made in this PR and/or \emph{why}. Such a description is helpful for reviewers and other developers to gain a quick understanding of the PR without touching the details and may reduce the possibility of the PR being ignored or rejected. However, developers sometimes neglect to write descriptions for PRs. For example, in our collected dataset with over 333K PRs, more than 34\% of the PR descriptions are empty. To alleviate this problem, we propose an approach to automatically generate PR descriptions based on the commit messages and the added source code comments in the PRs. We regard this problem as a text summarization problem and solve it using a novel sequence-to-sequence model.
To cope with out-of-vocabulary words in software artifacts and bridge the gap between the training loss function of the sequence-to-sequence model and the evaluation metric ROUGE, which has been shown to correspond to human evaluation, we integrate the pointer generator and directly optimize for ROUGE using reinforcement learning and a special loss function. We build a dataset with over 41K PRs and evaluate our approach on this dataset through ROUGE and a human evaluation. Our evaluation results show that our approach outperforms two baselines by significant margins.
\end{abstract}

 \section{Introduction}\label{sec:intro}

The pull-based development model~\cite{barr2012cohesive} is popular on modern collaborative coding platforms, e.g., GitHub~\cite{gousios2014exploratory, gousios2016work, website:github}. It eases developers' contributions to a project.
In this model, a developer does not need to have access to the central repository to contribute to a project.
She only needs to fork the central repository (i.e., create a personal clone), make changes (e.g., fix a bug or implement a feature) independently in the personal clone and submit the changes to the central repository through a pull request (from hereon, PR).
Usually, the PR will be tested by continuous integration services and reviewed by core team members (or reviewers) before being merged into the central repository~\cite{yu2015wait}.

A PR consists of one or more interrelated commits. To create a PR on GitHub, a developer needs to provide a title and can add a free-form text (i.e., a PR description) to describe further \emph{what} changes are made and/or \emph{why} they are needed.
PR descriptions can help reviewers gain a quick and adequate understanding of PRs without digging into details and may reduce the possibility of PRs being ignored or rejected~\cite{gousios2016work, fan2018early}.
In addition, PR descriptions can help software maintenance and program comprehension tasks~\cite{yu2014reviewer,fan2018early}.

However, PR descriptions are sometimes neglected by developers. 
For example, in our dataset which contains 333,001 PRs collected from 1K engineered Java projects on GitHub, over 34\% of the PR descriptions are empty.
To alleviate this problem, we propose an approach to automatically generate PR descriptions from the commits submitted with the corresponding PRs. 
Our approach can be used to generate PR descriptions to replace existing empty ones and can also assist developers in writing PR descriptions when creating PRs.

Some tools have been proposed to automatically generate descriptions for software changes, e.g., generate commit messages~\cite{buse2010automatically, linares2015changescribe, jiang2017automatically, liu2018neural} and release notes~\cite{moreno2014automatic, moreno2017arena}.
Commits, PRs and releases can be regarded as software changes occurring at different granularity.
Distinct from commit messages which only describe one commit, PR descriptions often need to summarize multiple related commits.
A release is a collection of plenty of commits and/or PRs.
Release notes are prepared for both developers and end users (people who use the libraries or apps), while PR descriptions' readers are usually solely developers. Hence they have different focuses and information structure.
Besides, the existing technique for release note generation~\cite{moreno2014automatic, moreno2017arena} does not explicitly summarize multiple interrelated commits.
It recovers links between commits and bug reports, and uses bug report titles as the summaries of corresponding commits.
In this work, we focus on explicitly summarizing the commits in a PR to generate its description, and we treat traceability as a separate problem.
Moreover, when developers document a change of some granularity (e.g., a PR), the documents of the smaller changes it contains (e.g., the commits in the PR) are usually available.
Therefore, the techniques of automatically documenting changes at different granularity are complementary rather than competing.
To the best of our knowledge, there is no prior work focusing on generating descriptions for PRs.

It is challenging to automatically generate a description for a single commit, not to mention a PR with multiple interrelated commits.
Fortunately, when a developer writes a PR description, the commit messages of the commits in this PR are usually available.
These valuable messages together with the patch of each commit shed light into the generation of PR descriptions.
As the first step of this task, this work aims to generate PR descriptions from the commit messages and the added source code comments in the PRs.

Given a PR, we regard the combination of its commit messages and the source code comments added in it as an ``article'' and its description as the summary of this ``article''.
The generation of PR descriptions is then regarded as a text summarization problem by us.
We have applied two commonly-used extractive text summarization methods to solve this problem but find their effectiveness is limited (described in Section~\ref{sec:RQ1}).
In this work, we propose a more effective approach for PR description generation.
Specifically. our approach builds upon the attentional encoder-decoder model~\cite{bahdanau2014neural}, which is an effective sequence-to-sequence model for text summarization.
It first learns how to write PR descriptions from existing PRs, and then can generate descriptions for new PRs.

There are two challenges which make a basic attentional encoder-decoder model not effective enough for PR description generation:

1) Out-of-vocabulary (OOV) words. 
Due to the developer-named identifiers (e.g., variable names), OOV words are widespread in software artifacts. 
However, the attentional encoder-decoder model can only produce words in a fixed vocabulary, hence cannot deal with OOV words.

2) The gap between the training loss function and the evaluation metric.
Since different sentences may convey similar meanings, researchers usually leverage a flexible discrete metric named ROUGE~\cite{lin2004rouge} to evaluate text summarization systems.
ROUGE allows for different word orders between a generated text and the ground truth, and correlates highly with human evaluation~\cite{lin2004rouge}.
However, the training objective of the attentional encoder-decoder model is minimizing a maximum-likelihood loss, which is strict and will penalize all literal differences between a generated text and the ground truth.
Due to this gap, the model minimizing the maximum-likelihood loss may not be the one with the best generation performance.

We observe that the OOV words in a PR description can often be found in the corresponding ``article''. 
Therefore we integrate the pointer generator~\cite{see2017get} in our approach to overcome the first challenge.
With this component, our approach can generate a word from either the fixed vocabulary or the input.
To deal with the second challenge, we need to find a way to optimize for ROUGE directly when training.
However, we cannot simply use ROUGE as the training objective since ROUGE scores are non-differentiable, i.e., the parameter gradients of our model cannot be calculated from them.
We solve this problem by using a reinforcement learning (RL) technique named self-critical sequence training (SCST)~\cite{rennie2017self}.
Based on SCST, we adopt a special loss function named RL loss~\cite{paulus2017deep} in our approach.
This loss is both related to ROUGE scores and differentiable, with which we can train a model and guide it to produce results that are more likely to be good for human evaluation.

As this is the first work on this topic, we use two extractive methods, i.e., LeadCM and LexRank~\cite{erkan2004lexrank}, as baselines.
Given a PR, LeadCM extracts its first few commit messages as the generated description, and LexRank selects and outputs salient sentences in the PR's ``article''.
To evaluate our proposed approach, we collected over 333K PRs from 1K engineered Java projects with the most merged PRs on GitHub, building a dataset with over 41K PRs after preprocessing.
We evaluate our approach on the dataset using ROUGE.
The evaluation results show that our approach outperforms the baselines in terms of ROUGE-1, ROUGE-2 and ROUGE-L by 11.6\%, 25.4\% and 12.2\%.
We also conduct a human evaluation to assess the quality of the generated PR descriptions, which shows that our approach performs significantly better than the baselines and can generate more high-quality PR descriptions.

In summary, our contributions are three-fold:

\begin{itemize}
    \item We propose a novel approach to generate descriptions for PRs from their commit messages and the code comments that are added. Our approach can cope with OOV words with the pointer generator and directly optimize for ROUGE, which has been shown to correspond to human evaluation, with the RL loss.
    \item We build a dataset with over 41K pull requests from GitHub for the PR description generation task.
    \item We evaluate our approach on the dataset using the ROUGE metric and a human evaluation. The evaluation results show that our approach outperforms two baselines by significant margins.
\end{itemize}

The remainder of this paper is organized as follows:
Section~\ref{sec:preliminary} describes the motivation, the usage scenarios and some background knowledge of our approach.
Section~\ref{sec:approach} elaborates our approach, including the pointer generator and the RL loss.
We describe our dataset in Section~\ref{sec:dataset} and present the procedures and results of our evaluation in Section~\ref{sec:evaluation}.
Section~\ref{sec:discussion} discusses situations where our approach performs badly and threats to validity.
After a brief review of related work in Section~\ref{sec:related_work}, we conclude this paper and point out potential future directions in Section~\ref{sec:conclusion}.
 \section{Motivation and Preliminary} \label{sec:preliminary}
In this section, we present the motivation and usage scenarios of our approach, formulate the problem of PR description generation, and introduce the attentional encoder-decoder model.

\subsection{Motivating Example}

\begin{table}[!t]
\centering
\caption{A pull request in the Pitest project}
\label{tab:moti_example}
\begin{tabular}{|p{0.47\textwidth}|}
\hline
\tabincell{p{0.47\textwidth}}{
\textbf{Description:}\\
Added an option to ignore failing tests from coverage, activated from maven plugin
}\\
\hline
\tabincell{p{0.47\textwidth}}{
\textbf{Commit 1:}\newline
\textbf{Commit Message:}
Added skipFailingTests option from maven plugin\newline
\textbf{Added Code Comments:} When set will ignore failing tests when computing coverage. Otherwise, the run will fail. If parseSurefireConfig is true, will be overridden from surefire configuration property testFailureIgnore\newline
\textbf{Commit 2:}\newline
\textbf{Commit Message:} Simplified surefire testFailureIgnore value retrieval\newline
\textbf{Added Code Comments:} N/A} \\
\hline
\end{tabular}
\end{table} 
Table~\ref{tab:moti_example} shows a motivating example of our approach, which is a PR in the Pitest project\footnote{https://github.com/hcoles/pitest/pull/528}.
We can see that the description describes the changes made, i.e., ``added an option'' and ``activated from maven plugin'', and the motivation, i.e., ``ignore failing tests from coverage'', of this PR.
This PR contains two commits.
One source code comment is added in Commit 1, and no code comment is added in Commit 2.
We can know the changes from the commit messages and the motivation from the added code comments in Commit 1.
This example indicates that we may be able to generate the description of a PR by summarizing its commit messages and the source code comments added in it.

\subsection{Usage Scenario}

Our approach aims to automatically generate descriptions for PRs based on their commit messages and the code comments that are added.
Its usage scenarios are as follows:

First of all, our approach can be used to generate PR descriptions to replace existing empty ones.
The generated descriptions may help reviewers and developers quickly capture PRs' key ideas without reading the detailed commits.
Such key ideas can be very helpful when reviewers and developers are making quick decisions, e.g., assigning a tag or estimating whether two PRs are related.
The generated descriptions may also be useful for software maintenance and program comprehension tasks.
For example, tools for PR reviewer recommendation can use the generated description as one of the features.

Our approach can also assist developers in writing PR descriptions.
If it takes several days to finish a PR, the developer may forget some important information in this PR when writing the description.
She may either ignore such information, which may affect the acceptance of the PR, or spend some time to check the detailed commits, which may decrease her productivity.
The description generated by our approach can remind the developer of the important information in the PR and assist her in writing a high-quality PR description.

\subsection{Problem Formulation}
Inspired by the motivating example, we regard the generation of a PR description as a text summarization task with the combination of the commit messages and the added code comments in the PR as the ``article'' and the PR description as the ``summary''.
Specifically, in this work, we treat the text summarization task as a sequence-to-sequence learning problem, where the source sequence is the ``article'' and the target sequence is the ``summary''. Therefore, the problem is formulated as follows: given a source sequence $\bm{w} = (w_1, w_2, ..., w_{|\bm{w}|})$ and a target sequence $\bm{y} = (y_1, y_2, ..., y_{|\bm{y}|})$, find a function f so that $f(\bm{w}) = \bm{y}$. $|\cdot|$ denotes the length of a sequence.

\subsection{Attentional Encoder-Decoder Model} \label{sec:encoder-decoder}

\begin{figure}[!t]
    \centering
    \includegraphics[width=0.48\textwidth]{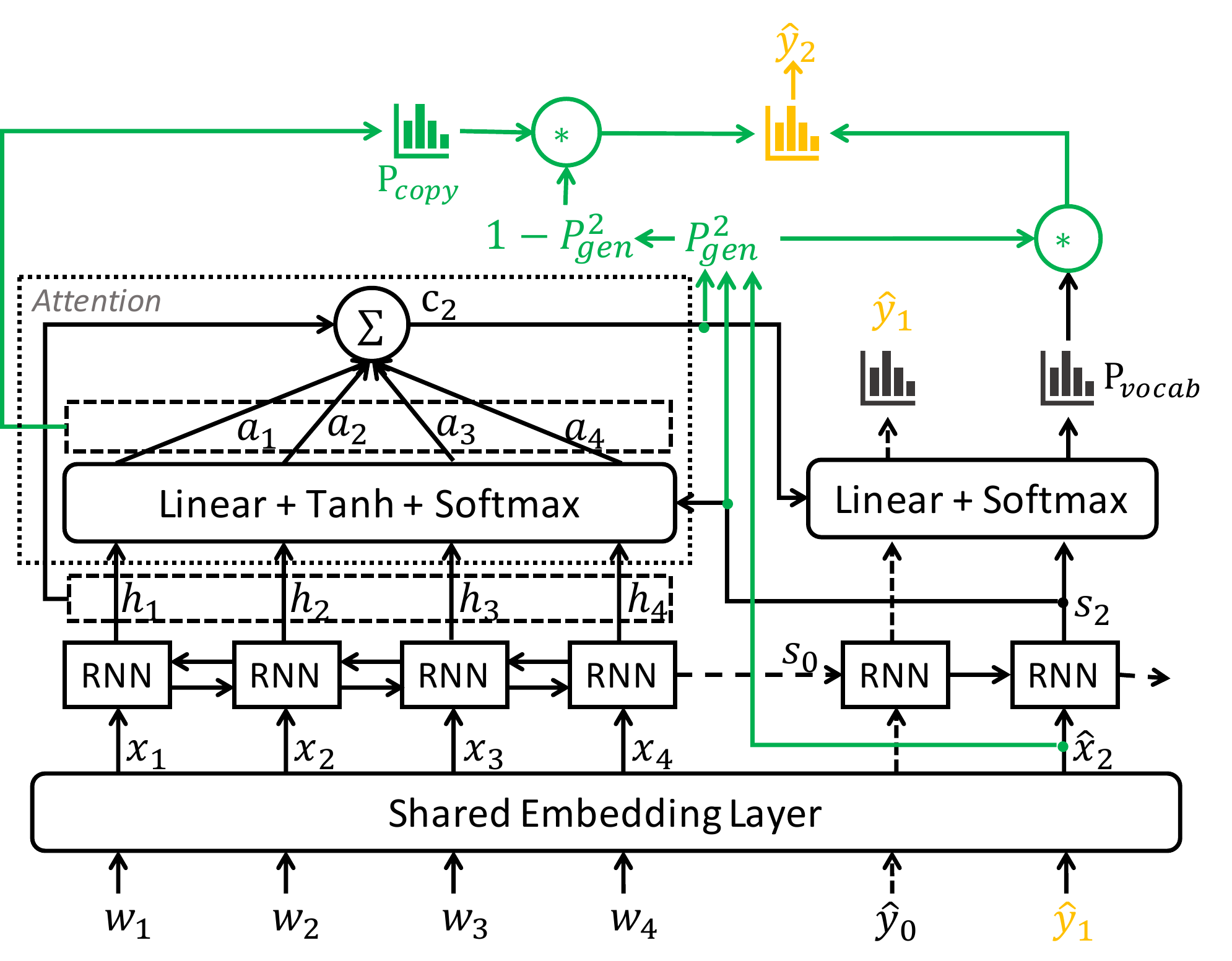}
    \caption{Attentional encoder-decoder model with pointer generator (Attn+PG)}
    \label{fig:pointer}
\end{figure}

Our approach builds upon the attentional encoder-decoder model~\cite{bahdanau2014neural} (from hereon, Attn), which is an effective model for sequence-to-sequence learning problems.
Attn's framework is depicted in black in Figure~\ref{fig:pointer}.
It uses two distinct recurrent neural networks (RNNs) as encoder and decoder, respectively.
The input of the encoder and the decoder is first mapped to word embeddings by a shared embedding layer. Given a source sequence $\bm{w}$, at time step $i$, the encoder calculates a hidden state $\bm{h_i}$ based on the word embedding $\bm{x_i}$ of $w_i$ and the previous hidden state $\bm{h_{i-1}}$ using its RNN.
The last hidden state $\bm{h_{|\bm{w}|}}$ is regarded as the intermediate representation of the source sequence and input to the decoder as the initial hidden state.

At decoding step $j$, the decoder computes a hidden state $\bm{s_j}$ from $\bm{\hat{x}_j}$, which is the embedding of the previous reference token when training or the previously generated token when testing.
Attn leverages the attention mechanism~\cite{bahdanau2014neural}, so the decoder also calculates a context vector $\bm{c_j}$, as follows:
$$e^{j}_i = \bm{v^e}tanh(\bm{W_hh_i} + \bm{W_ss_j} + \bm{b_e})$$
\begin{equation}
    \label{eq:attn_dist}
    \bm{a^j} = softmax(\bm{e^j})
\end{equation}
$$\bm{c_j} = \sum_{i}^{|\bm{x}|}\bm{a^j_{i}h_i}$$
where $\bm{W_h}$, $\bm{W_s}$ and $\bm{b_e}$ are learnable parameters and $e^{j}_i$ is the score of $x_i$ at decoding step $j$.
$\bm{a^j}$ is the attention distribution, which informs the decoder of the importance of each encoding step.
$\bm{c_j}$ is computed as the weighted sum of all encoder hidden states, which can be regarded as the representation of the source sequence at decoding step $j$.

Then, $\bm{c_j}$ is concatenated with decoder hidden state $\bm{s_j}$ to produce the vocabulary distribution $P_{vocab}$:
$$P_{vocab} = softmax(\bm{V'}(\bm{V}[\bm{s_j}, \bm{c_j}] + \bm{b}) + \bm{b'})$$
where $\bm{V'}$, $\bm{V}$, $\bm{b}$ and $\bm{b'}$ are learnable parameters. 
$P_{vocab}$ is used to decide which token in the vocabulary should be output at the current decoding step. It also provides the conditional probability of generating the $j_{th}$ reference token $y_j$, which is:
\begin{equation}
    \label{eq:word_prob}
    p(y_j|\hat{y}_0, \dots, \hat{y}_{j-1}, \bm{w}) = P_{vocab}(y_j)
\end{equation}
where $\bm{\hat{y}}$ is the input of the decoder.

At each training iteration, the optimization objective of Attn is to minimize the negative log-likelihood of the reference sequence, as follows:
\begin{equation}
    \label{eq:ml_loss}
    loss_{ml} = -\frac{1}{|\bm{y}|}\sum_{j=1}^{|\bm{y}|}\log p(y_j|\hat{y}_0, \dots, \hat{y}_{j-1}, \bm{w})
\end{equation}
 \section{Approach} \label{sec:approach}

This section elaborates our approach, including the pointer generator and the RL loss.

\subsection{Pointer Generator} \label{sec:pointer}

As described in Section~\ref{sec:encoder-decoder}, Attn produces tokens by selecting from a fixed vocabulary.
However, out-of-vocabulary (OOV) words are ubiquitous in software artifacts due to developer-named identifiers, such as variable names and file names.
Attn alone cannot cope with such OOV words, and hence its performance is limited.
We observe that the OOV words in PR descriptions usually appear in the corresponding source sequences. So we integrate the pointer generator~\cite{see2017get} in our approach to solve this problem. With this component, our approach can either select a token from the fixed vocabulary or copy one from the source sequence at each decoding step.

The structure of Attn with the pointer generator (Attn+PG) is presented in Figure~\ref{fig:pointer}, where the pointer generator is highlighted in green.
The switch between selection and copy is softly controlled by the \emph{generation probability}, which is calculated from the word embedding $\bm{\hat{x}_j}$ of the decoder input $\hat{y}_{j-1}$, the decoder state $\bm{s_j}$ and the context vector $\bm{c_j}$ at time step $j$:
$$p_{gen}^j = \sigma(\bm{w_c^Tc_j} + \bm{w_s^Ts_j} + \bm{w_x^T\hat{x}_j} + b_{gen})$$
where $\bm{w_c}$, $\bm{w_s}$, $\bm{w_x}$ and $b_{gen}$ are learnable parameters, and $\sigma$ is the sigmoid function.
$p_{gen}^j$ measures the probability that the $j_{th}$ output of the decoder is generated from the fixed vocabulary, and the probability of copy is hence $1-p_{gen}^j$. The conditional probability of producing the $j_{th}$ reference token (Equation~\ref{eq:word_prob}) is then modified as:
\begin{equation}
    \label{eq:new_word_eq}
    p(y_j|\hat{y}_0, \dots, \hat{y}_{j-1}, \bm{w}) = p_{gen}^jP_{vocab}(y_j) + (1-p_{gen}^j)P_{copy}(y_j)
\end{equation}
where $P_{copy}(y_j)$ is the probability of copying $y_j$ from the source sequence $\bm{w}$, and is computed from the attention distribution $\bm{a^j}$ (Equation~\ref{eq:attn_dist}), as follows:
$$P_{copy}(y_j) = \sum_{i:w_i=y_j}a_i^j$$

We can see that when $y_j$ is an OOV word, $P_{vocab}(y_j)$ is zero, but if $y_j$ appears in the source sequence, our approach can still generate it through $P_{copy}(y_j)$.
In this way, our approach alleviates the problem of OOV words and still holds the capability of producing novel words from the vocabulary.
In addition, the training loss is still calculated using Equation~\ref{eq:ml_loss}, but the conditional probability $p(y_j|\hat{y}_0, \dots, \hat{y}_{j-1}, \bm{w})$ is computed by Equation~\ref{eq:new_word_eq} now.
 
\subsection{RL Loss}
As described in Section~\ref{sec:encoder-decoder}, Attn uses the negative log likelihood loss to guide the training process. 
This loss function is strict and will penalize any literal difference between generated sequences and the ground truth.
For example, if the ground truth is ``the cat sat on the mat'' but the decoder produces ``on the mat sat the cat'', the loss will be high since the two sentences only literally match at ``the''.
Therefore, researchers do not use this loss function to measure the performance of text summarization systems. 
Instead, they usually use a flexible discrete evaluation metric named ROUGE (see Section~\ref{sec:metric}), which can tolerate generated sentences with different word orders from the ground truth and has been shown to correlate highly with human evaluation~\cite{lin2004rouge}.
The gap between the training loss function and ROUGE may result in the model with the least loss not being the one producing the best PR descriptions.

We can bridge this gap by directly optimizing for ROUGE when training.
However, ROUGE scores are non-differentiable, which means the parameter gradients of our model cannot be calculated only from ROUGE scores.
Hence we cannot directly use ROUGE as the loss function.
Recently, it has been shown that reinforcement learning (RL) techniques can be incorporated to enable direct optimization of discrete evaluation metrics~\cite{ranzato2015sequence, rennie2017self, paulus2017deep}.
Our approach also leverages an RL technique named self-critical sequence training (SCST)~\cite{rennie2017self} and adopts a special loss function named RL loss~\cite{paulus2017deep} to solve this problem.

We can cast the generation of PR descriptions using RL terminology.
The decoder is the ``agent'', which interacts with the ``environment'' (the encoder's output and the decoder's input).
At each decoding step, the ``action'' of the ``agent'' is to predict an output token according to a ``policy'' $\pi_{\theta}$ with parameters $\theta$. Actually, the neural network of the decoder defines $\pi_{\theta}$ and $\theta$ is its parameters.
Once finished generating a sequence $\hat{y}$, the ``agent'' will observe a ``reward'', defined as follows:
$$r(\bm{\hat{y}}) = g(\bm{\hat{y}}, \bm{y})$$
where $\bm{y}$ is the ground truth of $\bm{\hat{y}}$ and $g$ is a function related to ROUGE. In this work, we defined $g$ as the ROUGE-L F1 score.
The training objective of the RL problem is minimizing the negative expected reward:
$$L(\theta) = -\mathbb{E}_{{\bm{y^s}}\sim \pi_{\theta}}[r(\bm{y^s})]$$

\begin{figure}[!t]
    \centering
    \includegraphics[width=0.45\textwidth]{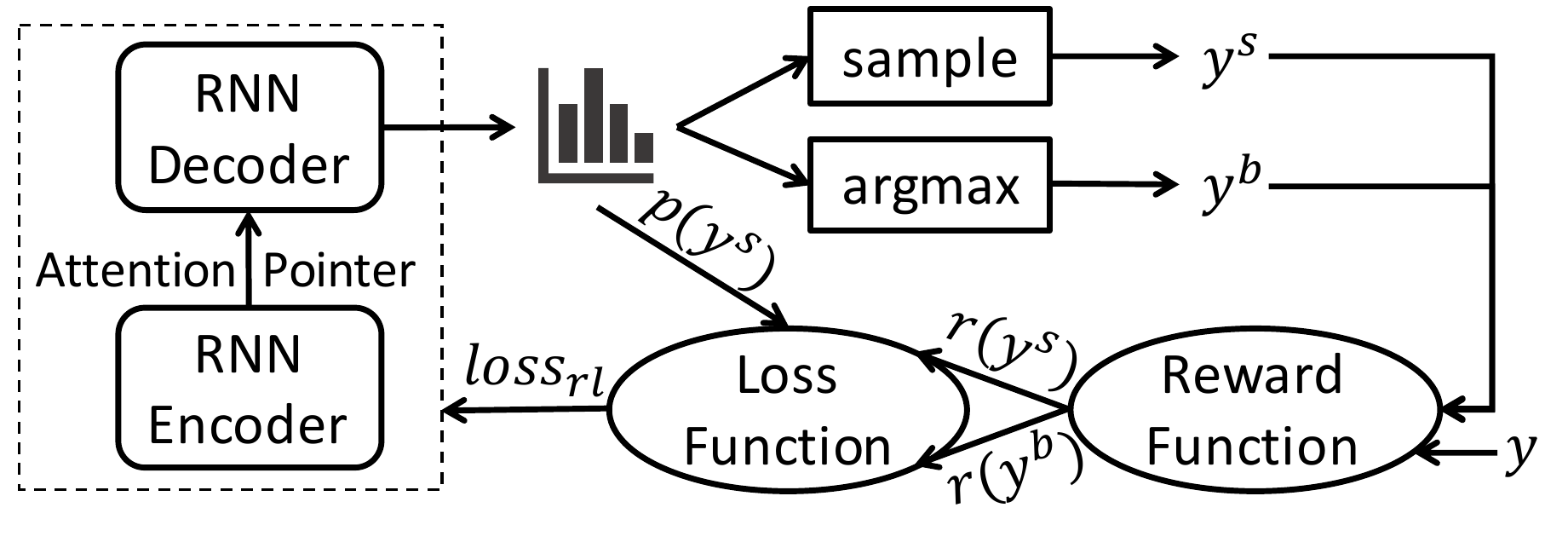}
    \caption{The RL loss}
    \label{fig:rl_loss}
\end{figure}

According to the SCST algorithm~\cite{rennie2017self}, the expected gradient of $L(\theta)$ can be computed as follows:
$$\nabla_{\theta}L(\theta) = -\mathbb{E}_{{\bm{y^s}}\sim \pi_{\theta}}[(r(\bm{y^s}) - r(\bm{y^b}))\nabla_{\theta}\log\pi_{\theta}(\bm{y^s})]$$
where $\bm{y^b}$ is a baseline sequence also generated from $\pi_{\theta}$. We obtain $\bm{y^b}$ through a greedy search. Specifically, we choose the token with the highest output probability (i.e., $p(y^b_j|y^b_0, \dots, y^b_{j-1}, \bm{w})$) at each decoding step to form $\bm{y^b}$.
In practice, the expectation can be approximated by a single Monte-Carlo sample $\bm{y^s}$ from $\pi_{\theta}$, which means:
\begin{equation}
    \label{eq:rl_object}
    \begin{split}
    &\nabla_{\theta}L(\theta)
         = -(r(\bm{y^s}) - r(\bm{y^b}))\nabla_{\theta}\log\pi_{\theta}(\bm{y^s})\\
        &= -(r(\bm{y^s}) - r(\bm{y^b}))\nabla_{\theta}\sum_{j=1}^{|\bm{y^s}|}\log p(y^s_j|y^s_0, \dots, y^s_{j-1}, \bm{w})
    \end{split}
\end{equation}
We define the RL loss of our model following Paulus et al.~\cite{paulus2017deep}, as follows:
\begin{equation}
    \label{eq:rl_loss}
    loss_{rl} = -(r(\bm{y^s}) - r(\bm{y^b}))\sum_{j=1}^{|\bm{y^s}|}\log p(y^s_j|y^s_0, \dots, y^s_{j-1}, \bm{w})
\end{equation}
$r(\bm{y^s})$ and $r(\bm{y^b})$ are non-differential and are regarded as constant values when calculating gradients. 
From Equation~\ref{eq:rl_object} and Equation~\ref{eq:rl_loss}, we can see that minimizing $loss_{rl}$ is equal to minimizing $L(\theta)$.
$loss_{rl}$ can also be viewed as the $loss_{ml}$ (Equation~\ref{eq:ml_loss}) weighted by a normalized reward. 
If the normalized reward, i.e., $r(\bm{y^s}) - r(\bm{y^b})$, is positive, i.e., the sequence sampled from our model is better than the baseline sequence, minimizing $loss_{rl}$ is equivalent to maximizing the likelihood of the sampled sequence, and vice versa.
The calculation process of the RL loss is shown in Figure~\ref{fig:rl_loss}.

In practice, only using $loss_{rl}$ can be detrimental to the readability of the generated texts~\cite{paulus2017deep}. We hence combine $loss_{ml}$ and $loss_{rl}$ to form a hybrid loss for training, as follows:
\begin{equation}
    \label{eq:hybrid_loss}
    loss = \gamma loss_{rl} + (1-\gamma) loss_{ml}
\end{equation}
 \section{Dataset} \label{sec:dataset}

\begin{figure*}[!t]
    \centering
    \includegraphics[width=0.88\textwidth]{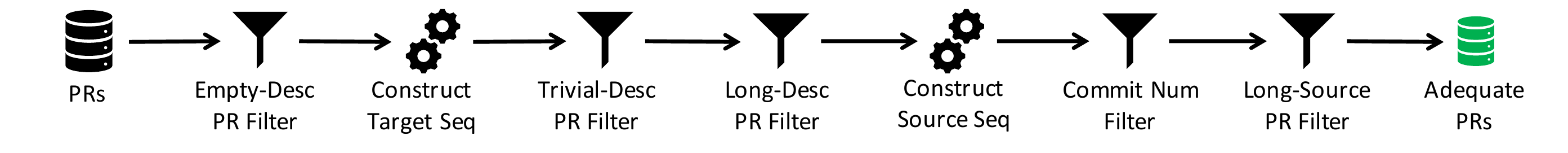}
    \caption{Procedure of filtering pull requests}
    \label{fig:prepro}
\end{figure*}

\begin{table*}[!t]
\centering
\ra{1.1}
\caption{Statistics of our collected pull requests}
\label{tab:dataset}
\begin{threeparttable}
\begin{tabular}{@{}ccccccccc@{}}
\toprule
\textbf{Type} & \tabincell{c}{\textbf{Empty-desc}\\ \textbf{PR}} & \tabincell{c}{\textbf{Trivial-desc}\\ \textbf{PR}} & \tabincell{c}{\textbf{Long-desc}\\ \textbf{PR}} & \tabincell{c}{\textbf{PR with only 1}\\ \textbf{valid commit}} & \tabincell{c}{\textbf{PR with \textgreater 20}\\ \textbf{valid commits}} & \tabincell{c}{\textbf{Long-source}\\ \textbf{PR}} & \textbf{Adequate PR} & \textbf{Total}\\
\textbf{Number} & 114,466 & 61,547 & 20,516 & 83,803 & 2,438 & 8,399 & 41,832 & 333,001\\
\bottomrule
\end{tabular}
\begin{tablenotes}
    \scriptsize
    \item *\textbf{Long-desc PR} and \textbf{Long-source PR} refer to the PRs for which the target sequence and the source sequence do not meet the length constraints, respectively.
\end{tablenotes}
\end{threeparttable}
\end{table*}
 
\subsection{Data Collection}
To collect PR data from GitHub, we first used the RepoReapers framework~\cite{munaiah2017curating} to select engineered software projects.
We obtained all 95,804 Java repositories that had been classified as containing engineered software projects by RepoReapers's Random Forest classification and retrieved the number of merged PRs for each repository.
22,700 of these repositories contained at least one merged PR.
We then sorted these 22,700 repositories in descending order of their number of merged PRs, downloading the data of merged PRs from the top 1,000 projects through GitHub's APIs. 
For each project, we collected at most the first 1,000 merged PRs returned by GitHub's search API. 
Since our approach takes commit messages and source code comments that are added as input and PR descriptions as output, given a PR, we retrieved its description and commit messages, parsed the patches of its commits and extracted the added comments in each patch.
In total, we collected 333,001 merged PRs from 1,000 engineered Java projects.

\subsection{Data Preprocessing} \label{sec:preprocessing}

We preprocessed the collected PRs according to the following processes:

\subsubsection{Preprocess text}\label{sec:text-prepro}
To filter out trivial and templated information in PRs, we leveraged the same procedure to preprocess the texts of PR descriptions, commit messages and source code comments.
Given a text, we first removed the HTML comments and the paragraphs starting with a headline named ``checklist'' from it through regular expressions, because the text in HTML comments and checklist paragraphs usually only describe the general procedure of finishing a PR, such as ``functionality works'' and ``passes all tests''.

Then we split the text into sentences using NLTK~\cite{website:nltk}, identifying and deleting the sentences with 1) url, 2) internal reference, e.g., ``\#123'', 3) signature, e.g., ``signed-off-by'', 4) emails, 5) `@name' and 6) markdown headlines, e.g., ``\#\# why'' through regular expressions. 
We filtered sentences with 1) and 2) since this work focuses on summarizing the changes in a PR and we regard recovering links between PRs and other software artifacts as a separate problem.
Besides, sentences with 3), 4), 5) and 6) usually do not describe the changes made in a PR and may bring in many OOV words. 

Next, we tokenized the text using NLTK.
Previous work has shown that NLTK outperforms other common NLP libraries in terms of tokenizing software documentation~\cite{al2017choosing}.
The tokens that only consist of 7 or more hexadecimal characters were considered as SHA1 hash digests and replaced with ``sha''. Similarly, version strings, e.g., ``1.2.3'', and numbers were converted into ``version'' and 0, respectively.

Finally, tokens with non-ASCII characters were removed and referred to as \emph{non-ASCII tokens}, and texts with more than 50\% \emph{non-ASCII tokens} were marked as ``non-ASCII''.

\subsubsection{Construct target sequence} 
The target sequence of a PR only consists of its description.
To construct it, we simply preprocessed the PR description using the general text preprocessing procedure mentioned above.
The PR descriptions which were marked as ``non-ASCII'' when preprocessing, contain less than 5 tokens or only consist of punctuation marks were removed and referred to as \emph{trivial desc}.

\subsubsection{Construct source sequence}
A PR's source sequence is constructed from the combination of its commit messages and the added code comments in it.
Specifically, we first listed the commits in this PR in ascending order of their creation time.
Then, for each commit, we extracted its commit message and the code comments that are added. The commit message was directly preprocessed using the general text preprocessing procedure.
As for the added comments, copyright comments, license comments, function signatures in Java docs (e.g., ``@param: param1'') and the comments with only punctuation marks were filtered.
The remaining comments were concatenated as a comment paragraph, which was then preprocessed using the general text preprocessing procedure.
If the preprocessed commit message or comment paragraph or both are not empty, we regard this commit as a \emph{valid commit}.
Finally, we concatenated all the preprocessed commit messages as the first paragraph of the source sequence and listed the comment paragraph of each commit as the following paragraphs. 
The commit messages were sorted according to the order of commits, i.e., ascending order of their creation time, and were separated by a special token ``[cm-sep]''.
The comment paragraphs were also listed in the order of commits, and all paragraphs were separated by ``[para-sep]''.

\subsubsection{Filter PRs}
With constructed source sequences and target sequences, PRs were filtered according to the procedure shown in Figure~\ref{fig:prepro}.
The PRs with empty descriptions were first removed and referred to as \emph{empty-desc PRs}.
If a PR description was a \emph{trivial desc} after preprocessing, the corresponding PR was also filtered and referred to as a \emph{trivial-desc PR}.
We deleted the PRs with less than 2 or more than 20 \emph{valid commits}, because we can directly use the only commit message as the description if a PR only contains one commit, and a PR with too many commits often aims to synchronize with another repository instead of being a contribution from a contributor. To reduce the training time of our approach, we also constrained the maximal length of the source sequence to be 400 and that of the target sequence to be 100. The PRs not satisfying these length constraints were hence filtered. 
After preprocessing, we collected 41,832 PRs. The statistics of the removed and the adequate PRs are presented in Table~\ref{tab:dataset}.
 \section{Evaluation} \label{sec:evaluation}
In this section, we first describe the evaluation metrics and the baselines. Then we present our research questions (RQs) and corresponding experiment results. Finally, we show the procedure and results of our human evaluation.

\subsection{Evaluation Metrics} \label{sec:metric}
We evaluate our approach with the ROUGE metric~\cite{lin2004rouge}, which has been shown to correlate highly with human assessments of summarized text quality~\cite{lin2004rouge}.
Specifically, we use ROUGE-N (N=1,2) and ROUGE-L, which are widely used to evaluate text summarization systems~\cite{see2017get, paulus2017deep}.

The recall, precision and F1 score for ROUGE-N are calculated as follows:
$$ R_{rouge\text{-}n} = \frac{\sum_{(gen, ref)\in S}\sum_{gram_n \in ref}Cnt_{gen}(gram_n)}{\sum_{(gen, ref)\in S}\sum_{gram_n \in ref}Cnt_{ref}(gram_n)} $$
$$ P_{rouge\text{-}n} = \frac{\sum_{(gen, ref)\in S}\sum_{gram_n \in ref}Cnt_{gen}(gram_n)}{\sum_{(gen, ref)\in S}\sum_{gram_n \in gen}Cnt_{gen}(gram_n)} $$
$$ F1_{rouge\text{-}n} = \frac{2R_{rouge\text{-}n}P_{rouge\text{-}n}}{R_{rouge\text{-}n} + P_{rouge\text{-}n}} $$
where $gen$, $ref$ and $S$ refer to a generated description, its reference description and the test set, $gram_n$ is an n-gram phrase and $Cnt_{gen}(gram_n)$ and $Cnt_{ref}(gram_n)$ refer to the occurrence number of $gram_n$ in $gen$ and $ref$, respectively. In summary, $R_{rouge\text{-}n}$ measures the percentage of the n-grams in reference descriptions that an approach can generate, and $P_{rouge\text{-}n}$ presents the percentage of ``correct'' n-grams (i.e., n-grams appearing in reference descriptions) in generated descriptions. $F1_{rouge\text{-}n}$ is a summary measure that combines both precision and recall.
The precision, recall and F1 score for ROUGE-L are similar with those for ROUGE-N, but instead of n-grams, they are calculated using the longest common subsequences between generated descriptions and reference descriptions~\cite{lin2004rouge}.
When comparing two approaches, we care more about F1 scores, since they balance precision and recall.

ROUGE is usually reported as a percentage value between 0 and 100. We obtained ROUGE scores using the pyrouge package~\cite{website:pyrouge} with Porter stemmer enabled.

\subsection{Baselines}
As this is the first work on PR description generation, we use two extractive baselines: LeadCM and LexRank.

\subsubsection{LeadCM} 
LeadCM is proposed by us for this task.
Given the source sequence of a PR, LeadCM outputs the first 25 tokens of the commit message paragraph as its generated description.
25 is the median length of the PR descriptions in our dataset.
According to the construction of the source sequence (described in Section~\ref{sec:preprocessing}), the generated description is actually the first few commit messages in this PR.
The hypothesis behind LeadCM is that developers may commit key changes, e.g., implementing a feature, first and make other less important changes, such as fixing typos, later, so the first few commit messages may summarize the core of a PR.

\subsubsection{LexRank}
LexRank~\cite{erkan2004lexrank} summarizes an article by calculating the relative sentence importance and selecting the most important sentences in the article as the generated summary.
It hypothesizes that a sentence is more salient to an article if it is similar to many sentences in this article.
The importance (or centrality) of each sentence in an article is computed by applying the PageRank algorithm~\cite{page1999pagerank} on the sentence similarity matrix of this article.
Given the source sequence of a PR, we first use the continuous LexRank method~\cite{erkan2004lexrank} to rank its sentences according to their importance. Then we concatenate these ranked sentences and keep the first 25 tokens as the output, just like LeadCM.

\subsection{Experiment Settings}
Similar to Jiang et al.~\cite{jiang2017automatically} and Hu et al.~\cite{hu2018deep}, we randomly select 10\% of the 41K PRs in our dataset for testing, 10\% for validation and the remaining 80\% for training.
Our approach uses 128-dimensional word embeddings.
The encoder is a single-layer bidirectional LSTM, the decoder is the same but unidirectional, and both of them use 256-dimensional hidden states.
Since our encoder and decoder share the embedding layer, we collect words from both the source sequences and the target sequences of the training set to build the vocabulary. The vocabulary size is set to 50K following See et al.~\cite{see2017get}.

At training time, we first train our model for 25,000 iterations only using the maximum-likelihood (ML) loss $loss_{ml}$, evaluating the model every 1,000 iterations with the validation set.
The best performing ML model is obtained after 12,000 iterations.
Then we continue training this best ML model with the hybrid loss $loss$ (defined in Equation~\ref{eq:hybrid_loss}) for another 28,000 iterations and also perform evaluation every 1,000 iterations.
We get the final best model after 22,000 iterations, i.e., 34,000 iterations in total.
We leverage Adam~\cite{kingma2014adam} with a batch size of 8 to train our models. 
As suggested by Paulus et al.~\cite{paulus2017deep}, we set the $\gamma$ in Equation~\ref{eq:hybrid_loss} to 0.9984, and the learning rate of Adam is set to 0.001 for the training with $loss_{ml}$ and 0.0001 for the training with $loss$.

When testing, we leverage beam search of width 4 to generate sequences.
We notice that there exist repeating phrases in some generated sequences and adopt a heuristic rule~\cite{paulus2017deep}, which ignores a candidate beam if its current generated token creates a duplicate trigram, at each decoding step to reduce such repetition.

Our replication package which contains our dataset, source code, trained model and test results is available online~\cite{website:replication, website:source_code}.

\subsection{RQ1: The Effectiveness of Our Approach} \label{sec:RQ1}

To investigate our approach's effectiveness, we evaluate our approach on our dataset in terms of ROUGE-1, ROUGE-2 and ROUGE-L, and compare it with the two baselines, i.e., LeadCM and LexRank.

\begin{table*}[!t]
\centering
\ra{1.1}
\caption{Comparisons of our approach (Attn+PG+RL) with each baseline in terms of ROUGE scores}
\label{tab:our_baseline}
\begin{threeparttable}
\begin{tabular}{@{}lcccccccccc@{}}
\toprule
\tworow{\textbf{Approach}} & \tworow{\textbf{Avg. length}} & \multicolumn{3}{c}{\textbf{ROUGE-1}} & \multicolumn{3}{c}{\textbf{ROUGE-2}} & \multicolumn{3}{c}{\textbf{ROUGE-L}}\\
\cmidrule(lr){3-5}\cmidrule(lr){6-8}\cmidrule(lr){9-11}
& & \textbf{Recall} & \textbf{Precision} & \textbf{F1 score} & \textbf{Recall} & \textbf{Precision} & \textbf{F1 score} & \textbf{Recall} & \textbf{Precision} & \textbf{F1 score}\\
\midrule
\textbf{LexRank} & 24.21 & 25.72 & 29.28 & 24.11 & 12.88 & 13.35 & 11.40 & 24.02 & 27.20 & 22.42\\
\textbf{LeadCM} & 24.37 & \textbf{33.16} & 36.31 & 30.61 & 20.29 & 20.28 & 17.85 & \textbf{31.42} & 34.17 & 28.89\\
\textbf{Attn+PG+RL} & 19.21 & 32.47 & \textbf{46.35} & \textbf{34.15} & \textbf{21.82} & \textbf{27.76} & \textbf{22.38} & 30.94 & \textbf{43.56} & \textbf{32.41}\\
\midrule
\textit{Attn+PG+RL vs LexRank} & \textit{-5.00} & \textit{26.27\%} & \textit{58.29\%} & \textbf{\textit{41.65\%}} & \textit{69.43\%} & \textit{107.92\%} & \textbf{\textit{96.33\%}} & \textit{28.84\%} & \textit{60.14\%} & \textbf{\textit{44.52\%}}\\
\textit{Attn+PG+RL vs LeadCM}  & \textit{-5.16} & \textit{-2.07\%} & \textit{27.65\%} & \textbf{\textit{11.57\%}} & \textit{7.51\%} & \textit{36.89\%} & \textbf{\textit{25.40\%}} & \textit{-1.51\%} & \textit{27.48\%} & \textbf{\textit{12.18\%}}\\
\bottomrule
\end{tabular}
\begin{tablenotes}
    \scriptsize
    \item *\textbf{Avg. length} refers to the average length of the generated descriptions by each approach.
\end{tablenotes}
\end{threeparttable}
\end{table*} 
The evaluation results are shown in Table~\ref{tab:our_baseline}, and our approach is referred to as Attn+PG+RL.
For text generation tasks, the F1 scores for ROUGE are typically between 0.2 to 0.4~\cite{wan2018improving, see2017get, paulus2017deep}.
We can see from Table~\ref{tab:our_baseline} that first our approach tends to generate shorter descriptions (19.21 tokens on average) than LexRank and LeadCM.
Second, it outperforms LexRank in terms of all metrics by large margins (from 32.35\% to 138.61\%).
Also, it obtains higher precision and F1 score than LeadCM for each ROUGE metric.
The improvements in terms of the three F1 scores are 3.54, 4.53 and 3.52 points, respectively.
These results indicate that compared to the two baselines, our approach can capture the key points of a PR more precisely.

\begin{table}[!t]
\centering
\caption{Test example 1}
\label{tab:example1}
\begin{tabular}{|p{0.47\textwidth}|}
\hline
\tabincell{p{0.47\textwidth}}{
\textbf{Source Sequence:}\\
initial tomcat 0 support . [cm-sep] tomcat 0 support in the s-ramp installer . [cm-sep] fixes for tomcat support . \newline
[para-sep]\\
eat the error and try the next option\\
eat the error and try the next option\\
this filter can be used to supply a source of credentials that can be used when logging in to the jcr repository ( modeshape ) . it uses the inbound request as the source of authentication .\\
constructor .\\
login with credentials set by some external force . this may be a servletfilter when running in a servlet container , or it may be null when running in a jaas compliant application server ( e.g . jboss ) .\\
note : when passing ' null ' , it forces modeshape to authenticate with either .} \\
\hline
\tabincell{p{0.47\textwidth}}{
\textbf{Reference:}\\
added support for tomcat 0 in s-ramp .
}\\
\hline
\tabincell{p{0.47\textwidth}}{
\textbf{LexRank:}\\
eat the error and try the next option
eat the error and try the next option
this filter can be used to supply
}\\
\hline
\tabincell{p{0.47\textwidth}}{
\textbf{LeadCM:}\\
initial tomcat 0 support .
tomcat 0 support in the s-ramp installer .
fixes for tomcat support .
}\\
\hline
\tabincell{p{0.47\textwidth}}{
\textbf{Attn+PG+RL:}\\
initial tomcat 0 support in the s-ramp installer .
}\\
\hline
\end{tabular}
\end{table} 
We also manually inspect our test results. Table~\ref{tab:example1} presents an example in the test set. 
According to our inspection, we argue that the better performance of our approach mainly comes from its two advantages:

1) Our approach can accurately identify important phrases or sentences in source sequences. 
It learns knowledge about which phrases are important for summarizing a PR from the training data, hence it is more intelligent.
For example, the PR in Table~\ref{tab:example1} contains 3 commit messages and multiple added code comments.
LexRank extracts the wrong sentences as output, performing worst. LeadCM outputs all commit messages without filtering or sorting, while our approach precisely identifies the salient phrase, i.e., ``tomcat 0 support in the s-ramp'' in the source sequence. 
Therefore, our approach outperforms the baselines.

2) Our approach is an abstractive method with the ability of dynamic generation.
LexRank and LeadCM generate descriptions by extracting important sentences and cannot rephrase extracted sentences, generate novel words or dynamically decide how many sentences/tokens to generate.
However, our approach can generate descriptions with different lengths based on the input.
Moreover, our approach can rephrase important sentences, as shown in Table~\ref{tab:example1}.
This advantage reduces the number of unimportant phrases produced by our approach and results in high precision.

We also notice that the recall of our approach for ROUGE-1 and ROUGE-L are slightly lower than that of LeadCM. This is because LeadCM always generates as many tokens as possible until it gets 25, while our approach tends to procedure short but accurate descriptions. 
Some phrases output by LeadCM may be trivial, but they may contain some tokens in the reference and hence improve the recall.
For example, in Table~\ref{tab:example1}, the third sentence output by LeadCM is not a salient sentence. However, it contains a ``for'', which also occurs in the reference and makes the recall of LeadCM higher than ours in this example.

In addition, it is a little surprising that the relatively complicated LexRank performs worse than the naive LeadCM. 
We argue that the reason is the hypothesis of LexRank, which is that a sentence similar to many sentences is more important, does not always hold in the source sequences.
For instance, in Table~\ref{tab:example1} there are two identical comments in the source sequence which are computed as the most important sentences by LexRank.
But they have no token in common with the reference.

\vspace{-0.05cm}
\begin{framed}
\vspace{-0.2cm}
In summary, our approach outperforms the two baselines in terms of ROUGE-1, ROUGE-2 and ROUGE-L, and can generate more accurate descriptions than the baselines.
\vspace{-0.2cm}
\end{framed}

\subsection{RQ2: The Effects of Main Components}

Our approach generates PR descriptions based on the attentional encoder-decoder model (Attn).
It integrates the pointer generator (PG) to deal with OOV words and adopts the RL loss to directly optimize for ROUGE when training.
We compare our approach, i.e., Attn+PG+RL, with Attn and Attn+PG in terms of ROUGE to understand the influence of the pointer generator and the RL loss.
Besides, since Attn is an effective and popular model for text generation tasks, we also regard it as an abstractive baseline for PR description generation and comparing our approach with it can further investigate the effectiveness of our approach.

\begin{table*}[!t]
\centering
\ra{1.1}
\caption{Effectiveness of each component in our approach}
\label{tab:our_comp}
\begin{threeparttable}
\begin{tabular}{@{}lcccccccccc@{}}
\toprule
\tworow{\textbf{Approach}} & \tworow{\textbf{Avg. length}} & \multicolumn{3}{c}{\textbf{ROUGE-1}} & \multicolumn{3}{c}{\textbf{ROUGE-2}} & \multicolumn{3}{c}{\textbf{ROUGE-L}}\\
\cmidrule(lr){3-5}\cmidrule(lr){6-8}\cmidrule(lr){9-11}
& & \textbf{Recall} & \textbf{Precision} & \textbf{F1 score} & \textbf{Recall} & \textbf{Precision} & \textbf{F1 score} & \textbf{Recall} & \textbf{Precision} & \textbf{F1 score}\\
\midrule
\textbf{Attn} & 13.95 & 19.68 & 36.23 & 22.92 & 11.32 & 18.13 & 12.74 & 18.90 & 34.54 & 21.95\\
\textbf{Attn+PG} & 14.02 & 27.68 & \textbf{47.22} & 31.27 & 19.32 & \textbf{28.66} & 21.15 & 26.48 & \textbf{44.61} & 29.82\\
\textbf{Attn+PG+RL} & 19.21 & \textbf{32.47} & 46.35 & \textbf{34.15} & \textbf{21.82} & 27.76 & \textbf{22.38} & \textbf{30.94} & 43.56 & \textbf{32.41}\\
\midrule
\textit{PG} & \textit{+0.07} & \textit{40.63\%} & \textit{30.31\%} & \textbf{\textit{36.47\%}} & \textit{70.73\%} & \textit{58.08\%} & \textbf{\textit{66.10\%}} & \textit{40.15\%} & \textit{29.15\%} & \textbf{\textit{35.87\%}}\\
\textit{RL}  & \textit{+5.19} & \textit{17.33\%} & \textit{-1.84\%} & \textbf{\textit{9.21\%}} & \textit{12.90\%} & \textit{-3.14\%} & \textbf{\textit{5.81\%}} & \textit{16.82\%} & \textit{-2.36\%} & \textbf{\textit{8.68\%}}\\
\bottomrule
\end{tabular}
\begin{tablenotes}
    \scriptsize
    \item *\textbf{Attn}, \textbf{PG} and \textbf{RL} refer to the attentional encoder-decoder model, the pointer generator and the RL loss, respectively.
\end{tablenotes}
\end{threeparttable}
\end{table*} \begin{table}[!t]
\centering
\caption{Test example 2}
\label{tab:example2}
\begin{tabular}{|p{0.47\textwidth}|}
\hline
\tabincell{p{0.47\textwidth}}{
\textbf{Source Sequence:}\newline
add some more links to fultextonline.\newline
fulltextonline and webpagearchived in parallel rather than choosing between one of them ) but this should be ok.\newline
- add when 6551ex starts with ' onl '\newline
- add when medium is rdvocab : 0\newline
- add resource ht018400499 to the test set\newline
- adjust test sets .\newline
[cm-sep]\newline
add rule to avoid redundant entries in fulltextonline and webpagearchived\newline
introduced with the last commit :\newline
we will now have fulltextonline and webpagearchived not in parallel but rather choose between them , with preferring webpagearchived .\newline
- adjust test sets .
}\\
\hline
\tabincell{p{0.47\textwidth}}{
\textbf{Reference:}\newline
we will now have fulltextonline and webpagearchived not in parallel but rather choose between them , with preferring webpagearchived .\newline
- add when 6551ex starts with ` onl '\newline
- add when medium is rdvocab : 0\newline
- adjust test sets
}\\
\hline
\tabincell{p{0.47\textwidth}}{
\textbf{Attn:}\newline
adapt and [UNK] in parallels .
}\\
\hline
\tabincell{p{0.47\textwidth}}{
\textbf{Attn+PG:}\newline
fulltextonline and webpagearchived in parallel rather than choosing between one of them ) but this should be ok.
}\\
\hline
\tabincell{p{0.47\textwidth}}{
\textbf{Attn+PG+RL:}\newline
fulltextonline and webpagearchived in parallel rather than choosing between one of them ) but this should be ok.\newline
- add when 6551ex starts with ` onl '\newline
- adjust test sets .
}\\
\hline
\end{tabular}
\end{table} 
The evaluation results are shown in Table~\ref{tab:our_comp}.
We can see that Attn+PG outperforms Attn in terms of all metrics by 29.15\% to 70.73\%, which means the pointer generator can effectively cope with OOV words and the generation of PR descriptions benefits a lot from it.
Table~\ref{tab:example2} presents one of our test results. We can see that ``fulltextonline'' and ``webpagearchived'' are two OOV words. Attn cannot handle them hence produces ``[UNK]'' instead, while both Attn+PG and Attn+PG+RL can generate the two words correctly.

Compared to Attn+PG, Attn+PG+RL performs better in terms of all recall and F1 score metrics by more than 5.8\%, but slightly worse in terms of all precision metrics. To figure out the reason, we inspect the descriptions generated by Attn+PG and Attn+PG+RL, and find that Attn+PG+RL tends to generate longer descriptions than Attn+PG in order to increase the RL reward, i.e., the F1-score for ROUGE-L. For example, Attn+PG+RL generates more relevant tokens than Attn+PG for the PR in Table~\ref{tab:example2}. On average, Attn+PG+RL produces 5.19 more tokens than Attn+PG, as shown in Table~\ref{tab:our_comp}. Therefore, the reduced precision of Attn+PG+RL can be regarded as the expense of the improved recall; and the gain in recall is higher than the loss in precision, which translates to high F1 scores.

\vspace{-0.05cm}
\begin{framed}
\vspace{-0.2cm}
In summary, our approach outperforms Attn and Attn+PG. The pointer generator and the RL loss are effective and helpful for boosting the effectiveness of our approach.
\vspace{-0.2cm}
\end{framed}

\subsection{Human Evaluation} \label{sec:human_eval}
We also conduct a human evaluation to investigate our approach's effectiveness.
We invite 6 human evaluators to assess the quality of the PR descriptions generated by our approach and the two baselines. All of them are Ph.D. students with 1-5 years of experience in Java programming.

\subsubsection{Procedure}
We randomly select 100 PRs from the test set and evenly divide them into two groups. Each group is evaluated by 3 different evaluators. For each PR, we show its source sequence and reference description followed by the three PR descriptions generated by our approach and the two baselines to the evaluators. The three generated descriptions are randomly ordered. Human evaluators also have no idea about how these approaches work, so they cannot figure out which description is generated by which approach. The evaluators are asked to assign a score from 0 to 7 to each generated description to measure the semantic similarity between the generated description and the reference.
The higher the score the closer is the corresponding PR description to the reference.
Evaluators are allowed to search related information, such as unfamiliar concepts, through the Internet.

\begin{figure}[!t]
    \centering
    \includegraphics[width=0.48\textwidth]{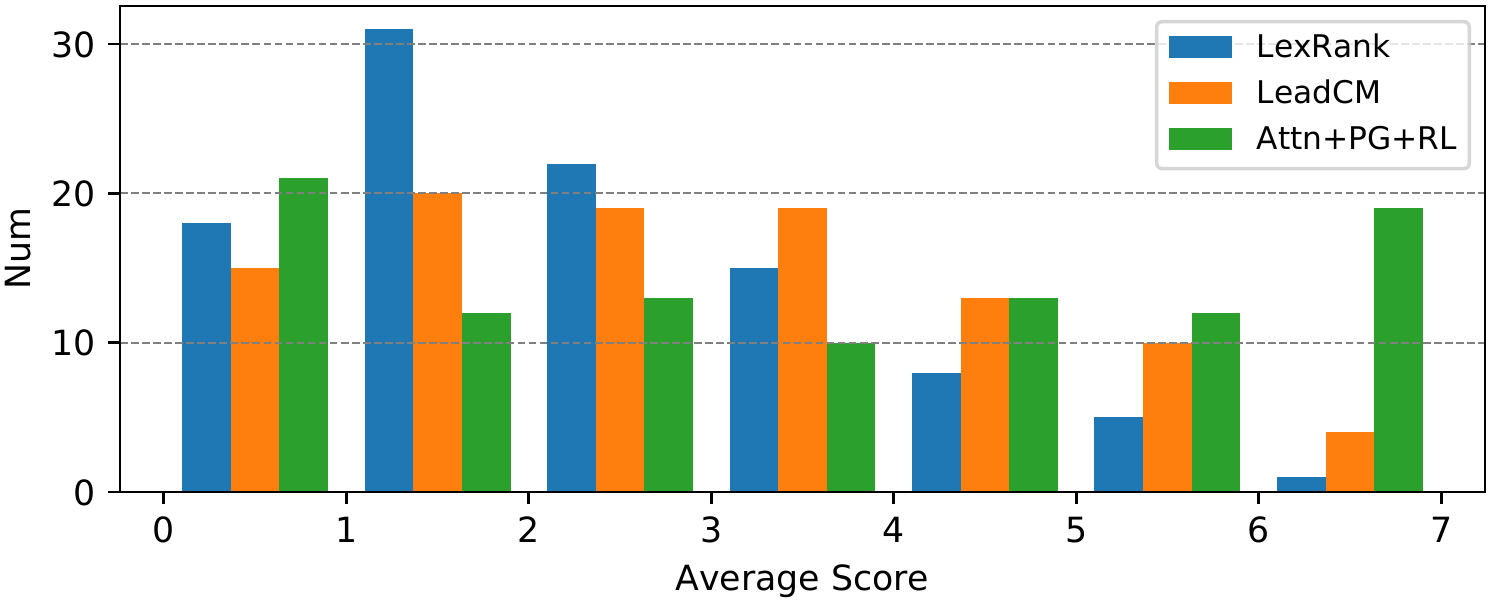}
    \caption{The distribution of the \emph{final scores} obtained from our human evaluation. Each bar presents the number of the average scores obtained by an approach that fall in a specific score interval. For example, the leftmost blue bar shows that 18 descriptions generated by LexRank obtain an average score between 0 to 1.}
    \label{fig:user_study}
\end{figure}

\subsubsection{Results}

Each PR description obtains three scores from three evaluators. We calculate the average score as its \emph{final score}.
The distribution of the \emph{final scores} is presented in Figure~\ref{fig:user_study}.
We can see that compared to the baselines, our approach produces more PR descriptions with high scores ($\geqslant 4$) and less with low scores ($< 4$).
Besides, our approach generates far more PR descriptions with the average score between 6 and 7.
But we also notice that our generated descriptions with the average score between 0 and 1 are a little more than those generated by the baselines.
The reason may be that our approach generates PR descriptions from scratch instead of directly extracting source sentences, and hence sometimes may fail to generate important words.

We  calculate the average scores of LexRank, LeadCM and Attn+PG+RL across all sampled PRs, which are 2.14, 2.73 and 3.27, respectively. Although the average score of our approach is still not perfect, our approach is the first step on this topic and can inspire follow-up work. We also conduct Wilcoxon signed-rank tests~\cite{wilcoxon1945individual} at the confidence level of 95\% considering the 100 \emph{final scores} for each approach. The p-values of our approach compared with LexRank and LeadCM are all less than 0.01, which means the improvements achieved by our approach are significant.

\vspace{-0.05cm}
\begin{framed}
\vspace{-0.2cm}
In summary, our human evaluation shows that our approach outperforms the baselines significantly, and can generate more high-quality PR descriptions.
\vspace{-0.2cm}
\end{framed}

\section{Discussion} \label{sec:discussion}
In this section, we discuss situations where our approach performs badly and threats to validity.

\begin{table}[!t]
\centering
\caption{Test example 3}
\label{tab:example3}
\begin{tabular}{|p{0.47\textwidth}|}
\hline
\tabincell{p{0.47\textwidth}}{
\textbf{Source Sequence:}\newline
improve performance of preprocess . [cm-sep] improve performance of preprocess , cleanup code . [cm-sep] improve performance of preprocess , cleanup code\newline
[para-sep]\newline
newcommand = newcommand.replaceall ( '\textbackslash\textbackslash\textbackslash\textbackslash ( [ \textasciicircum\textbackslash\textbackslash\textbackslash\textbackslash ( ] *\textbackslash\textbackslash\textbackslash\textbackslash\ ) ' , empty ) ;\newline
newcommand = newcommand.replaceall ( ' ; . * ' , empty ) ;\newline
build up the decimal formatter .\newline
build up the regular expression .\newline
return command.replaceall ( `\textbackslash\textbackslash\textbackslash\textbackslash' , empty ) ; .\newline
[para-sep]\newline
only build the decimal formatter if the truncation length has changed .\newline
build up the decimal formatter .\newline
build up the regular expression .
}\\
\hline
\tabincell{p{0.47\textwidth}}{
\textbf{Reference:}\newline
when opening large files , the preprocessing takes a very long time . by compiling the regular expressions once rather than using the string.replaceall methods , we can save significant time .\newline
on my laptop , i opened a gcode file with sha lines ( laser image raster )\newline
with my changes - 16155ms\newline
master - 56178ms
}\\
\hline
\tabincell{p{0.47\textwidth}}{
\textbf{Attn+PG+RL:}\newline
improve performance of preprocess , cleanup code .
}\\
\hline
\end{tabular}
\end{table} 

\subsection{Where Does Our Approach Perform Badly} \label{sec:bad_situations}
We carefully inspect the PRs where our approach does not obtain good ROUGE-L F1 scores.
We find that our approach usually performs badly if the reference description mainly presents the information that cannot be found in the source sequence. We find three types of such information: 1) context information, including the motivation and the test results of a PR; 2) implementation details; 3) subjective sentences, i.e., sentences describing personal feelings, plans, etc.

Table~\ref{tab:example3} presents a typical example.
We can see that the reference description of this PR contains three sentences, which respectively describe the motivation, the implementation details and the test results of the PR, and such information does not appear in the source sequence.
The description generated by our approach has little in common with the reference description, hence it gets low ROUGE scores.

As for subjective sentences, we find a test PR for which the reference description is ``we did make awesome, shall we land it''. This description is a subjective sentence without describing the changes made in the PR, while our approach tries to summarize the PR by generating ``refactored common pieces out into httpobject.''.
Upon our inspection, the description produced by our method correctly captures the meaning of the PR, while the reference description is not helpful for understanding the changes well.

Sometimes, our approach also fails to capture key phrases in the source sequence and consequently performs badly. But this situation is less common than the one mentioned above. Our evaluation results in Section~\ref{sec:RQ1} also show that our approach can better capture key phrases than the two baselines.

\subsection{Threats to Validity} \label{sec:threats}

One threat to validity is that our dataset was built only from Java projects, which may not be representative of all programming languages. 
However, Java is a popular programming language. Besides, our approach takes commit messages and source code comments as input hence can also be applied to projects of other programming languages.

Another threat to validity is that the non-summary information, such as signatures and subjective sentences, in PR descriptions may affect the effectiveness of our approach. 
PR descriptions are free-form text, and we cannot guarantee their quality and content. 
We mitigate this threat by using a set of heuristic rules to filter out non-summary information when preprocessing.
But it is hard to distill the patterns of all non-summary information. Since this work focuses on learning to generate PR descriptions from existing PRs, further improvements on data preprocessing are more suitable for future work.

There is also a threat related to our human evaluation.
We cannot guarantee that each score assigned to every PR description is fair.
To mitigate this threat, each sampled PR is evaluated by 3 human evaluators, and we use the average score of the 3 evaluators as the final score.

In addition, our baseline approaches produce summaries of length 25 tokens unless there are fewer than 25 tokens available in the source sequence. This may result in incomplete sentences in the output of these approaches, which may have negatively affected the corresponding ratings by human evaluators. However, this threat does not affect the ROUGE scores which also confirm the superiority of our approach.

\section{Related Work} \label{sec:related_work}
This section discusses the related studies on documenting software changes, understanding pull requests, and summarizing and documenting other software artifacts.

\subsection{Documenting Software Changes}
Commits, PRs and releases are software changes of different granularity. Some tools have been proposed to document commits based on diverse inputs automatically~\cite{buse2010automatically, le2014dynamic, cortes2014automatically, linares2015changescribe, shen2016automatic, rastkar2013did, jiang2017automatically, liu2018neural}.
For example, Buse and Weimer proposed \textsc{DeltaDoc}~\cite{buse2010automatically}, a technique that can summarize a commit by first using symbolic execution and path predicate analysis to generate the behavioral difference and then applying some heuristic transformations to generate a natural language description.
Similarly, Cortes et al. built ChangeScribe~\cite{cortes2014automatically}, a tool which first identifies the stereotype of a commit from the abstract syntax trees before and after the commit and then generates a descriptive commit message using pre-defined filters and templates.
Rastkar and Murphy~\cite{rastkar2013did} proposed an approach to generate the motivation of a commit by extracting motivational sentences from its relevant documents.
Jiang et al.~\cite{jiang2017automatically} adopted an attentional encoder-decoder model to generate commit messages from \texttt{diffs}.
Liu et al.~\cite{liu2018neural} proposed an information-retrieval-based method to generate commit messages for new \texttt{diffs} by reusing proper existing commit messages.

Researchers have also explored the automatic generation of release notes~\cite{abebe2016empirical, moreno2014automatic, moreno2017arena}.
For instance, Abebe et al.~\cite{abebe2016empirical} identified six types of information contained in release notes and leveraged machine learning techniques to decide whether an issue should be mentioned in release notes.
Moreno et al. proposed a tool named ARENA~\cite{moreno2014automatic, moreno2017arena}, which first extracts and summarizes each commit in a release, and then uses manually defined templates to organize these summaries with their related information in the issue tracker to generate release notes.

Commit messages, PR descriptions, and release notes document software changes occurring at different granularity~\cite{fu2015automated, yan2016automatically, gousios2016work, abebe2016empirical}.
As described in Section~\ref{sec:intro}, PR descriptions often need to summarize several related commits when compared to commit messages, and are different from release notes in terms of target audiences and information structure.
Moreover, since the documents of small-granularity changes (e.g., the commits in a PR) are usually available when developers document a large-granularity change (e.g., the PR), the techniques for generating commit messages and release notes are complementary rather than competing with our approach.

\subsection{Understanding Pull Requests}
Many empirical studies were focusing on understanding pull requests (PRs) and the pull-based development.
Some of them focused on analyzing which factors affect the PR evaluation~\cite{rahman2014insight,tsay2014influence, yu2015wait}. 
For example, Rahman and Roy~\cite{rahman2014insight} investigated how the discussion texts, project-specific information (e.g., project maturity) and developer-specific information (e.g., experience) of PRs affect their acceptance.
Tsay et al.~\cite{tsay2014influence} found that both technical and social factors influence the acceptance of PRs.

Some other studies aimed to understand how the pull-based development works~\cite{gousios2014exploratory, gousios2015work, gousios2016work}.
For instance, Gousios et al.~\cite{gousios2014exploratory} analyzed the popularity of the pull-based development model, characterized the lifecycle of PRs and also explored which factors influence the merge decision and delay of a PR. 
In their following work, Gousios et al.~\cite{gousios2015work, gousios2016work} conducted large-scale surveys to study how integrators (people who are responsible for integrating PRs) and contributors collaborate in the pull-based development model.
They highlighted the challenges faced by integrators, such as maintaining projects' quality and prioritizing external contributions, and the challenges faced by contributors like unawareness of project status and poor responsiveness from integrators.

Prior work also proposed many techniques to deal with the challenges developers face when using the pull-based development~\cite{yu2014reviewer, van2015automatically, zanjani2016automatically, jiang2017should, fan2018early}.
For example, Veen et al.~\cite{van2015automatically} proposed PRioritizer, a tool which prioritizes PRs based on machine learning techniques and multiple extracted features such as the number of discussion comments.
To reduce the response time of PRs, Yu et al.~\cite{yu2014reviewer} proposed an approach to automatically recommend reviewers for a PR based on its title, description, and the social relations of developers.
These studies motivate our work to generate PR descriptions to facilitate downstream tasks.

\subsection{Summarizing and Documenting Other Software Artifacts}

Besides software changes, researchers have studied the automatic summarization of other software artifacts, such as source code~\cite{sridhara2010towards, haiduc2010use, moreno2013automatic, wong2013autocomment, mcburney2014automatic, iyer2016summarizing, hu2018deep, hu2018summarizing, wan2018improving, hu2019deep}, bug reports~\cite{rastkar2010summarizing, rastkar2014automatic, mani2012ausum, lotufo2015modelling}, app reviews~\cite{di2016would}, developer discussions~\cite{uddin2017automatic, huang2018automating, viviani2019locating} and development activity~\cite{treude2015summarizing}. Concerning source code, some techniques have been proposed to summarize source code based on program analysis and manually-defined templates~\cite{sridhara2010towards, moreno2013automatic, mcburney2014automatic}, information retrieval~\cite{haiduc2010use, wong2013autocomment}, and learning-based methods~\cite{iyer2016summarizing, hu2018deep, wan2018improving}.
Some of them also use encoder-decoder models.
For example, Hu et al.~\cite{hu2018deep} proposed a framework which leveraged an attentional encoder-decoder model to generate comments for Java methods.
Wan et al.~\cite{wan2018improving} proposed a novel encoder-decoder model with a hybrid encoder and a reinforcement-learning-based decoder to generate code comments.
They used the actor-critic algorithm~\cite{konda2000actor} with an extra neural network as the critic. Different from Wan et al.'s work, our approach uses the SCST algorithm~\cite{rennie2017self}, which is based on the REINFORCE algorithm~\cite{williams1992simple} and does not require extra networks. Besides, we do not directly leverage RL to decode but only to compute a special loss for better training.

As for bug reports, previous work focused on identifying and extracting important sentences from bug reports as their summaries.
For example, Rastkar et al.~\cite{rastkar2010summarizing, rastkar2014automatic} trained a conversion-based summarizer using a bug report corpus to identify important sentences automatically. 
Mani et al.~\cite{mani2012ausum} and Lotufo et al.~\cite{lotufo2015modelling} proposed unsupervised approaches based on noise reducer~\cite{mani2012ausum} or heuristic rules~\cite{lotufo2015modelling} to perform bug report summarization.
Different from their work, this work aims to generate PR descriptions from commit messages and source code comments that are added using an abstractive method.

 \section{Conclusion and Future Work} \label{sec:conclusion}

In this paper, we aim to automatically generate descriptions for pull requests from their commit messages and the source code comments that are added.
We formulate this problem as a sequence-to-sequence learning problem and point out two challenges of this problem, i.e., out-of-vocabulary words and the gap between the training loss function of sequence-to-sequence models and the discrete evaluation metric ROUGE, which has been shown to correspond to human evaluation~\cite{lin2004rouge}.
We propose a novel encoder-decoder model to solve this problem.
To handle out-of-vocabulary words, our approach adopts the pointer generator to learn to copy words from the source sequences.
As for the second challenge, our approach incorporates a reinforcement learning technique and adopts a special loss function to optimize for ROUGE directly.
Comprehensive experiments on a dataset with over 41K pull requests and a human evaluation show that our approach outperforms two competitive baselines.

In the future, we plan to further investigate the usefulness of our approach by using it to generate summaries for the pull requests without descriptions. We also plan to improve our approach by involving additional related software artifacts as input. For example, by taking diff files and relevant bug reports as input, our approach may be able to infer the implementation details and the motivation of a PR.
 
\section*{Acknowledgment}
This research was partially supported by the National Key Research and Development Program of China (2018YFB1003904), NSFC Program (No. 61972339) and the Australian Research Council’s Discovery Early Career Researcher Award (DECRA) funding scheme (DE180100153).

\balance
\bibliographystyle{IEEEtran}
\bibliography{reference}

\end{document}